\documentclass[a4paper,dvipsnames,svgnames,table]{article}

\usepackage[english]{babel}
\usepackage[utf8x]{inputenc}
\usepackage[T1]{fontenc}
\usepackage{graphicx}

\usepackage[a4paper,top=3cm,bottom=2cm,left=2cm,right=2cm]{geometry}

\usepackage{multicol}
\graphicspath{ {images/} }

\usepackage{amsmath}
\usepackage{amsfonts}
\usepackage{amssymb}
\usepackage{bbm}
\usepackage[colorinlistoftodos]{todonotes}
\usepackage[colorlinks=true, allcolors=blue]{hyperref}
\usepackage{microtype}
\usepackage{authblk}

\usepackage[font=small,labelfont=bf]{caption}

\title{Allostery and cooperativity in multimeric proteins: bond-to-bond propensities in ATCase} 

\author[1,3]{Maxwell Hodges}
\author[2,3]{Mauricio Barahona}
\author[1,3]{Sophia N. Yaliraki}
\affil[1]{\textit{Department of Chemistry, Imperial College London, South Kensington Campus, London~SW7~2AZ, United Kingdom}}
\affil[2]{\textit{Department of Mathematics, Imperial College London, South Kensington Campus, London~SW7 2AZ, United Kingdom}}
\affil[3]{\textit{Institute of Chemical Biology, Imperial College London, South Kensington Campus, London~SW7 2AZ, United Kingdom}}

\begin{document}

\maketitle

\begin{abstract}
Aspartate carbamoyltransferase (ATCase) is a large dodecameric enzyme with six active sites that exhibits allostery: its catalytic rate is modulated  by the binding of various substrates at distal points from the active sites. 
 A recently developed method, bond-to-bond propensity analysis, has proven capable of predicting allosteric sites in a wide range of proteins using an energy-weighted atomistic graph obtained from the protein structure and given knowledge only of the location of the active site. 
Bond-to-bond propensity establishes if energy fluctuations at given bonds have significant effects on any other bond in the protein, by considering their propagation through the protein graph.
In this work, we use bond-to-bond propensity analysis to study different aspects of ATCase activity using three different protein structures and sources of fluctuations.  
First, we predict key residues and bonds involved in the transition between inactive (T) and active (R) states of ATCase by analysing allosteric substrate binding as a source of energy perturbations in the protein graph. Our computational results also indicate that the effect of multiple allosteric binding is non linear: a switching effect is observed after a particular number and arrangement of substrates is bound suggesting a form of long range communication between the distantly arranged allosteric sites.  
Second, cooperativity is explored by considering a bisubstrate analogue as the source of energy fluctuations at the active site, also leading to the identification of  highly significant residues to the T $\leftrightarrow$ R transition that enhance cooperativity across active sites. 
Finally, the inactive (T) structure is shown to exhibit a strong, non linear communication between the allosteric sites and the interface between catalytic subunits, rather than the active site. 
Bond-to-bond propensity thus offers an alternative route to explain allosteric and cooperative effects in terms of detailed atomistic changes to individual bonds within the protein, rather than through phenomenological, global thermodynamic arguments.

\end{abstract}

\section{Introduction}

Much has been written about allostery, the process through which binding of a molecule distal to the active site of a protein causes an attenuation or an enhancement in the catalytic rate of that protein~\cite{Nussinov2016, Guo2016, Ribeiro2016}. Yet the physical mechanisims underpinning this effect are still not well understood at the microscopic level, thus limiting the potential for chemical design and intervention.  
Most of the previous work on allostery has focussed on thermodynamic models linking changes in catalytic rates to modifications in the conformation of the protein.  Such an outlook led to the traditional models of allostery: the Monod-Wyman-Changeaux (MWC) model~\cite{Monod1963}, whereby binding of allosteric substrates causes a concerted conformational shift of the protein subunits towards the active state, and the Koshland-Nemethy-Filmer (KNF) model~\cite{koshland1966comparison}, which proposed that binding of an allosteric substrate to a subunit drives the latter towards the active state and the overall transition to the full active state is sequential.  More recently, Hilser and coworkers proposed the ensemble allosteric model (EAM)~\cite{Motlagh2014}, which rationalises allosteric outcomes according to the effect of the substrates on the entire conformational ensemble of the protein. Furthermore, there is a growing appreciation of the role of dynamics in allostery~\cite{Guo2016, Volkman2001}, including the role that entropy plays in the modelling of the energy landscape, which has led to the design of protein switches~\cite{Choi2015}.

\begin{figure}[!htb]
\centering
\includegraphics[width=0.8\textwidth]{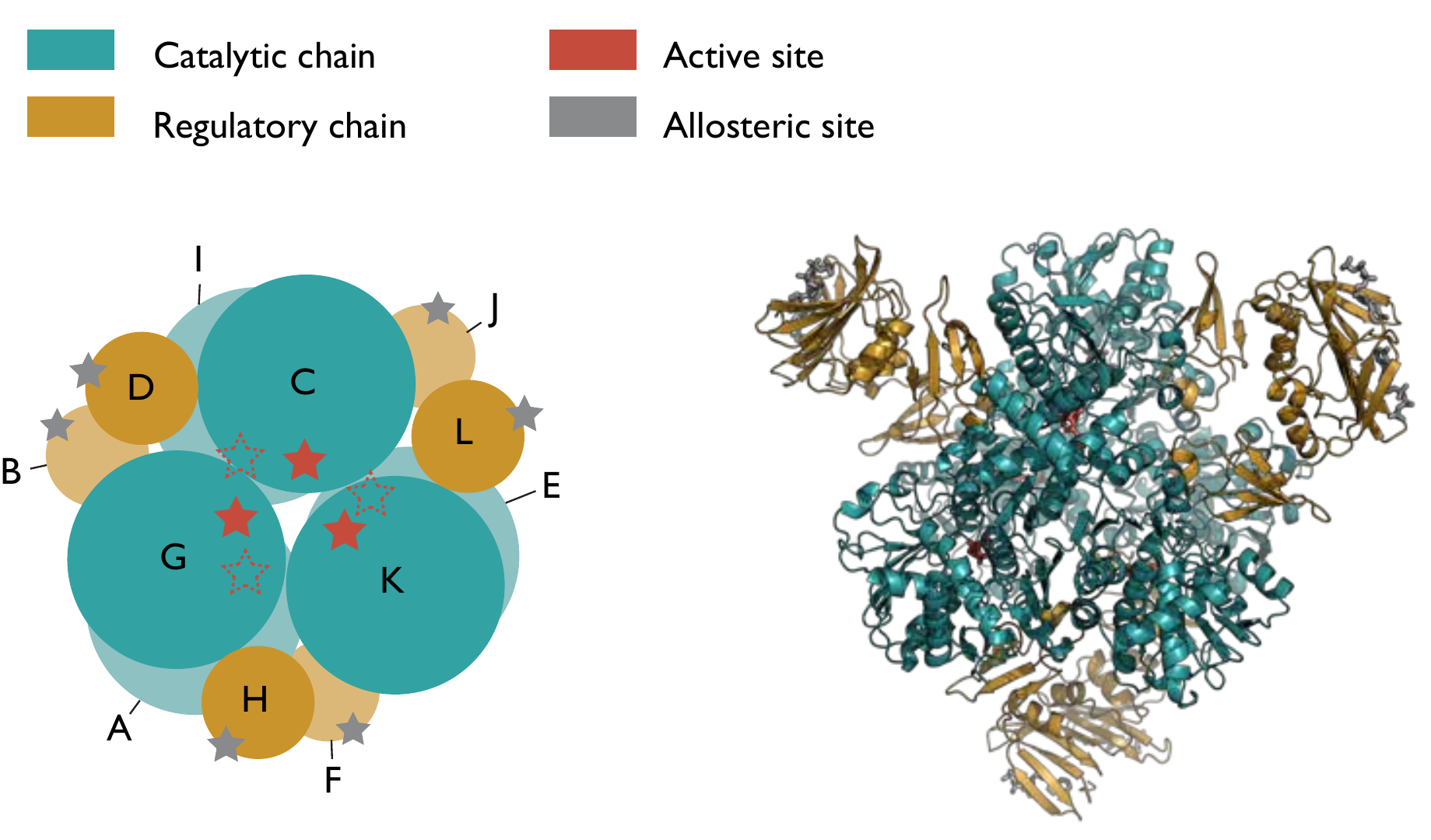}
\caption{\bf   ATCase comprises of  six catalytic and six regulatory subunits, shown in green and gold respectively, with more than 43000 atoms.  PALA (red) is a bisubstrate analogue of the reaction substrates (carbamoyl phosphate and aspartate) and sits in the active site, while ATP and CTP bind to the regulatory subunits and are shown in silver.} 
\label{ATCase_cartoon}
\end{figure}

Whilst thermodynamic models of allostery provide understanding of the equilibrium effects of substrate binding, they are unable to provide a detailed description of how a signal is transmitted between the allosteric binding site and the active site at the microscopic scale.  The so called \emph{structural view} of allostery posits that some form of propagation pathway between the allosteric and active sites must exist as a condition for allostery, though the existence of such a pathway does not imply allosteric behaviour by itself.  Tsai and Nussinov~\cite{Tsai2014} argue that both a structural and a thermodynamic component is required for a complete description of allostery.   Leitner notes~\cite{Leitner2008} that there are two alternate descriptions of this energy transfer:  the traversing of energy from one residue to another along structural pathways (often utilised in discussions of energy dispersion after photoexcitation~\cite{Koyama2006404, doi:10.1021/jp034558f}) or energy transfer between the normal modes of the protein~\cite{doi:10.1021/jp9813286, Kholodenko200071}.  
A significant advance regarding the possibility of structural pathways in allosteric proteins was the work of Lockless and Ranganathan, who used a statistical approach on evolutionary data to demonstrate coupling between residues in the PDK family of proteins situated between a binding and a distal site~\cite{Lockless1999}.  
However, the study of energy propagation in proteins is far from trivial experimentally, particularly on large proteins.  Dyer \textit{et al}~\cite{Li2014} used ultrafast infrared spectroscopy to examine the flow of energy within albumin and found that the flow was ballistic and anisotropic rather than diffusive, supporting the idea that structural pathways exist within proteins that allow for efficient energy transfer between coupled sites.  One of the challenges involved in experimental studies of allostery is that structural changes upon ligand binding can be subtle.  Falk \textit{et al}~\cite{Falk2016} exploited the extreme sensitivity of NMR chemical shifts to small structural changes to address this problem.  By using mutational studies to create singly bound thymidylate synthase dimers, they demonstrated that binding of the first allosteric effector \emph{primes} the enzyme for the binding of the second effector, such that both effectors are required for the allosteric response.  

Due to the challenges inherent to the experimental studies of allostery, a wide range of computational methods have been developed to model allosteric behaviour~\cite{Ribeiro2016}.  Perhaps the most common methods, particularly for larger proteins, are those that employ elastic network models (ENMs).  In general, ENMs model a protein as a system of balls and springs, where usually each ball represents a $C_{\alpha}$ atom, such that the protein is coarse-grained at the residue level.  The potential energy function for the springs is assumed to be a quadratic function and the overall dynamics is then described via normal mode analysis (NMA)~\cite{Lopez-Blanco2016}.  Improvements to the spring force constants have been found by direct comparison with molecular dynamics (MD) simulations~\cite{Ahmed2010}, whilst MD simulations themselves have been shown to provide insight into communication pathways in proteins~\cite{Ghosh2007}.  However, MD simulations suffer from high computational cost when applied to fully atomistic descriptions, and, moreover, it is often difficult to understand the coupling of dynamics across the relevant scales. This issue is particularly problematic for the large range transitions that are often involved in allosteric proteins.  

Network based approaches have become increasingly common, typically at the residue level of description.  Examples include studying changes in residue contacts upon allosteric substrate binding~\cite{Daily2008}, identifying residues involved in shortest paths\cite{DelSol2009} or using MD trajectories to build up so called protein energy networks (PENs)~\cite{Ribeiro2015}.  An important insight provided by Ribeiro and Ortiz is that when residue motion correlations are used to create the network, statistical errors render the results less accurate than when interaction energies are used, as a result of the high sensitivity of the signalling pathways to the network topology~\cite{Ribeiro2014}.  Recently, strain analysis of residues was applied to crystal structures~\cite{Mitchell2016}, which showed that sites of shear strain correspond to binding sites on the protein, thus providing physical insight into how allosteric substrates may transfer energy through a protein structure.  Network-theoretic machine learning tools have also been applied to fully atomistic protein graphs~\cite{Delmotte2011, Amor2014} demonstrating that a wealth of information can be obtained from static structures, avoiding the time consuming calculations often involved in molecular dynamics or Monte Carlo approaches. 
 
Bond-to-bond propensity analysis is a recently developed method~\cite{Amor2016}, which has previously been used to predict allosteric sites in a wide set of proteins through knowledge only of the active site of those proteins. This graph-theoretical method  was initially introduced to study flow redistribution in electrical networks and can be thought of as a graph-theoretical analogue of a Green's function in edge space~\cite{Schaub2014}.  A number of features of the method stand out.  Firstly, it uses a fully atomistic, energy-weighted graph description of the protein~\cite{Delmotte2011, Amor2014}, and hence it does not rely on any coarse-graining techniques to reduce the complexity of the protein structure.  Secondly, despite maintaining atomistic detail, the method remains computationally efficient; the calculations are carried out in almost linear time with respect to the number of edges as a result of recent work in algorithmic matrix theory~\cite{Spielman2004, Kelner}.  Furthermore, in contrast to most network approaches, bond-to-bond propensity analysis is focused on the interactions between \emph{edges} in a network  (i.e., the \emph{bonds} in a biomolecular system), and can thus reveal the significance of individual bonds in response to energy propagation. It is through bonds that energy fluctuations and flow propagate in a protein and this appears to be the key link between the mathematical basis of the method and the physical processes in the protein, as shown by the robustness of the method in identifying the allosteric sites for a set of 20 benchmark proteins~\cite{Amor2016}. 

The success of bond-to-bond propensity analysis in predicting allosteric sites motivates our work here, in which we study the `reverse' process; that is, we use bound allosteric substrates as the source of perturbation so as to replicate how fluctuations spreads physically throughout the protein structure. This allows us to identify the residues and bonds that are particularly crucial to energy transport.  By comparing the 'reverse' and 'forward' processes on both active and inactive states of the protein, we aim to explain how the different energy propagation processes may affect the transition between the two states and the allosteric effect altering the catalytic rate of the protein. 

Aspartate carbamoyltransferase (ATCase) is a dodecameric enzyme with six catalytic and six regulatory subunits 
(Fig.~\ref{ATCase_cartoon}). ATCase catalyses the reaction between L-aspartate and carbamoyl phosphate to form N-carbamyl-L-aspartate, the first step in the pyrimidine biosynthesis pathway.  ATCase has two main states: an inactive \emph{tense} state (T state) and an active \emph{relaxed} state (R state)~\cite{Lipscomb2012}. The transition between the T and R states produces a major change in shape and symmetry in this protein complex.  ATCase has been extensively studied for over 50 years~\cite{Gerhart2014, Changeux2013} as a paradigmatic example of two important phenomena in protein activity (for an extensive review see Ref.~\cite{Kantrowitz2012}).  Firstly, ATCase exhibits cooperativity, whereby binding of the reaction substrates at the active site results in a shift in the equilibrium towards the active R state.  Experimentally, this shift can be achieved by addition of the bisubstrate analogue N-(phosphonoacetyl)-L-aspartate (PALA)~\cite{Jin1999}.  Secondly, ATC exhibits allostery, with allosteric effectors  ATP and CTP as part of a negative feedback mechanism that controls the levels of pyrimidine in the cell. ATP increases the catalytic rate whilst CTP decreases it, and both substrates bind at the same position on the protein, though in slightly different configurations.  
Whilst both ATP and CTP  bind to both forms and cause slight changes in the quaternary structure, the binding of ATP to the inactive T state and CTP to the active R state is not sufficient to cause a population shift to the opposite state~\cite{Kantrowitz2012}.  

Although ATCase has been extensively studied, the microscopic mechanism that underpins allostery and cooperativity in this large protein complex is still not fully understood. In particular, the role that each of the molecules play in the different subunits in bringing about the allosteric transition remains elusive.  Bond-to-bond propensity analsysis is well suited to probing these questions as it is computationally efficient to model the {\it whole} complex at the atomistic level , thus allowing us to gauge the individual effect each perturbation has on any other  bond in the multimer.  ATCase is a large protein, consisting of 43,134 atoms and to the best of the authors' knowledge, the dodecameric ATCase structure has not previously been studied using a fully atomistic method. 

\section{Material and Methods}

\subsection{Structural data}

The three X-ray crystal structures of ATCase used in this work were downloaded from the Protein Data Bank (PDB)~\cite{Berman2000}.  
We studied two active state structures: 4KGV,  the R state bound to ATP (obtained at 1.2\AA~resolution~\cite{Cockrell2013}); and 1D09,  the unligated active state (resolved at 2.1\AA~\cite{Jin1999}).
We also used one inactive structure: 5AT1, the T state bound to CTP (obtained at 2.6\AA~resolution~\cite{Stevens1990}).

\subsection{Construction of the atomistic protein graph}

The initial step in the method is the conversion of the 3-dimensional coordinates of the atoms of the protein from the PDB file to a \emph{weighted graph}; that is, a collection of nodes (here representing the atoms) and edges (bonds, interactions) that link them.  The \emph{weight} of an edge between two nodes corresponds to the interaction energy of a bond or weak interaction obtained through atomic potentials. The procedure for the atomistic graph construction has been described in detail in Refs.~\cite{Delmotte2011, Amor2014, Amor2016} and here we summarise the main features.  The crystal structures typically do not contain hydrogen atoms and so the program Reduce (v.3.23)~\cite{Word19991735} is used to add these.  Following this, the software FIRST~\cite{PhysRevLett.75.4051, jacobs2000computer} is used to identify covalent bonds and non-covalent interactions (hydrogen bonds using a threshold of 0.01 $\text{kcal mol}^{-1}$, hydrophobic tethers with a  distance cutoff of 8\AA \ and salt bridges). Covalent bonds  are weighted using standard bond energies~\cite{HuheeyJamesEandKeiterEllenAandKeiterRichardLandMedhi2006}; hydrogen bonds  according to the potential in Ref.~\cite{Dahiyat1997}; and hydrophobic interactions using the potential developed by Lin \textit{et al}~\cite{Lin2007}.  Finally, electrostatic interactions of ions and ligands recorded by the LINK entries are accounted for using a standard Coulomb potential with atomic charges for the residues assigned using the OPLS-AA force field~\cite{jorgensen1988opls}.

\subsection{Bond-to-bond propensity}

The formulation of \emph{bond-to-bond propensity} was presented in detail in Ref.~\cite{Amor2016} and thus is briefly summarised here.  The key matrix that defines bond-to-bond propensities is $M$, the $m \times m$ bond-to-bond transfer matrix, where $m$ is the number of edges or bonds.  The element $M_{ji}$ describes how a perturbation at bond $i$ is transmitted to bond $j$ via a propagation that includes the entire graph structure~\cite{Schaub2014}.  $M$ is shown to be given by 
\begin{equation}
M = \frac{1}{2} W B^T L^{\dagger} B
\label{propensity}
\end{equation}
where $B$ is the $n \times m$ incidence matrix for the graph with $n$ nodes and $m$ edges and $L^{\dagger}$ is the pseudo-inverse of the weighted Laplacian matrix $L$, which governs the diffusion dynamics on the energy-weighted graph~\cite{lambiotte2014random}.  The weighted Laplacian is given by:
\begin{equation}
L=\begin{cases}
    -\omega_{ij}, & \text{$i\neq j$}.\\
    \sum_{j} \omega_{ij}, & \text{$i = j$},
  \end{cases}
\end{equation}
where $\omega_{ij}$ corresponds to the interaction energy between atoms $i$ and $j$.  More compactly, the Laplacian can be rewritten as $L=B W B^T$ where $W=\text{diag}(\omega_{ij})$ is a $m \times m$ diagonal matrix that contains the energy of interactions of all edges on the diagonal.

To evaluate the effect of perturbations from a group of bonds $b'$ (e.g., belonging to a ligand) on another bond $b$ we select the corresponding columns of the matrix $M$ and compute the sum of the absolute values in the $b^{th}$ row of the selected columns:
\begin{equation}
\Pi_b^\text{ raw} = \sum_{b' \in \, \text{ligand}} |M_{bb'}|
\end{equation}
where $b'$ includes all the weak bonds between the protein and the source (i.e., the ligand). 

The \textit{bond propensity} is then defined as: 
\begin{equation}
\label{bond_score}
\Pi_b = \dfrac{\Pi_b^\text{ raw}}{\sum_{b}\Pi_b^\text{ raw}},
\end{equation}
which is normalised by the total propensity score of all the bonds in the system.

The results presented in this paper are often in the form of the \textit{residue propensity}, which is calculated by summing over the normalised bond propensities of the bonds belonging to the residue $R$:
\begin{equation}
\Pi_R = \sum_{b \in R} \Pi_b. 
\label{residue_score}
\end{equation}

\subsection{Quantile regression}
  
As is physically expected, the propensity of a bond within the protein decays away from the perturbation source. To detect significant effects in the protein structure, we need to compare bond propensities at a similar distance from the source, thus taking into account the expected effect of distance. This is achieved using \textit{conditional quantile regression} (QR)~\cite{Koenker1978}, which allows us to identify high propensity bonds at the tail of the highly non-normal distribution~\cite{Amor2016}.

The distance of a bond from the perturbation source is taken to be the minimum distance between that bond $b$ and any of the bonds of the chosen source residues: 
\begin{equation}
d_b = \underset{b' \in \text{source bonds}}{\text{min}} |\mathbf{x}_b - \mathbf{x}_{b'}|,
\end{equation}  
where $\mathbf{x}_b$ holds the cartesian coordinates of the midpoint of bond $b$.  Because propensity scores are seen to generally fall away exponentially with distance, the logarithm of the propensity is used to generate the parameters in the QR minimisation problem:
\begin{equation}
\hat{\boldsymbol{\beta}}_b^\text{prot}(p) = \underset{(\beta_{b,0}, \ \beta_{b,1})}{\text{argmin}} \sum_{b}^{\text{protein}} \rho_{p} \left( \log(\Pi_b) - (\beta_{b,0} + \beta_{b,1}d_{b}) \right)   
\end{equation}
where 
\begin{equation}
\rho_{p}(y) = \left \vert y\ (p - \mathbbm{1}(y < 0)) \right \vert
\end{equation}
is the QR loss function to be minimised for each quantile $p$ and $\mathbbm{1}$ denotes the indicator function.  
The result of this optimisation is the model $\hat{\boldsymbol{\beta}}^\text{prot} =(\hat{\beta}^\text{prot}_{b,0} (p), \hat{\beta}^\text{prot}_{b,1} (p))$ that describes the quantiles of the propensities for all bonds in the protein.  

The \textit{bond quantile score} can then be calculated for each bond in the protein by finding the quantile $\rho_p$ such that:
\begin{equation}
p_{b} = \underset{p \in [0,1]}{\text{argmin}} \left \vert \log(\Pi_{b}) - (\hat{\beta}_{b,0}^\text{prot}(p) + \hat{\beta}_{b,1}^\text{prot}(p)d_{b})  \right \vert
\end{equation}
for bond $b$ with propensity $\Pi_b$ at a distance $d_b$ from the source bonds.  The corresponding \textit{residue quantile score} ($p_R$) is similarly defined, instead using residue propensities and the minimum distance between the atoms of each residue and those of the source bonds:
\begin{equation}
\hat{\boldsymbol{\beta}}_R^\text{prot}(p) = \underset{(\beta_{R,0}, \ \beta_{R,1})}{\text{argmin}} \sum_{R}^{\text{protein}} \rho_{p} \left( \log(\Pi_R) - (\beta_{R,0} + \beta_{R,1}d_{R}) \right)   
\end{equation}
and
\begin{equation}
p_{R} = \underset{p \in [0,1]}{\text{argmin}} \left \vert \log(\Pi_{R}) - (\hat{\beta}_{R,0}^\text{prot}(p) + \hat{\beta}_{R,1}^\text{prot}(p)d_{R})  \right \vert
\label{residue_QR}
\end{equation}

We can then use this bond quantile score (and its corresponding residue analogue $p_R$) to establish which bonds (and residues) have significantly propensities once the distance effect has been regressed out. 

Our quantile regression calculations make use of the R library \emph{quantreg} written by R. Koenker~\cite{koenker2015quantreg}.\\

The datasets generated during and analysed during the current study are available from the corresponding author on request.

\section{Results}

\subsection{Allostery: Active R State with ATP sources at allosteric sites}

\subsubsection{Identification of key residues and bonds under full allosteric occupation}

ATP is an allosteric activator of ATCase, able to increase the activity of the enzyme by 180\% at a 2mM concentration~\cite{Kantrowitz2012}.  ATP does not affect the maximal rate of the enzyme; instead, it induces a shift from the inactive T state to the active R state. From a global thermodynamic perspective, the MWC and EAM models would suggest this shift is caused by a preferential stabilisation of the active R state over the inactive T state, whilst the KNF model would attribute the shift to the binding of ATP to the inactive state driving it towards the active state. 

Here, we instead focus on the changes in energy flow within the protein as a result of different substrate binding.

We analyse the full atomistic graph obtained from the crystal structure of ATCase in the active R state (4KGV) with six allosteric binding sites for ATP.  
 
Bond-to-bond propensity analysis is then used to identify significant bonds and residues under allosteric perturbation sources, as well as the effect of changing the number of ATP sources, as a proxy for ATP concentration.

\begin{figure}[!htb]
\centering
\includegraphics[width=0.8\textwidth]{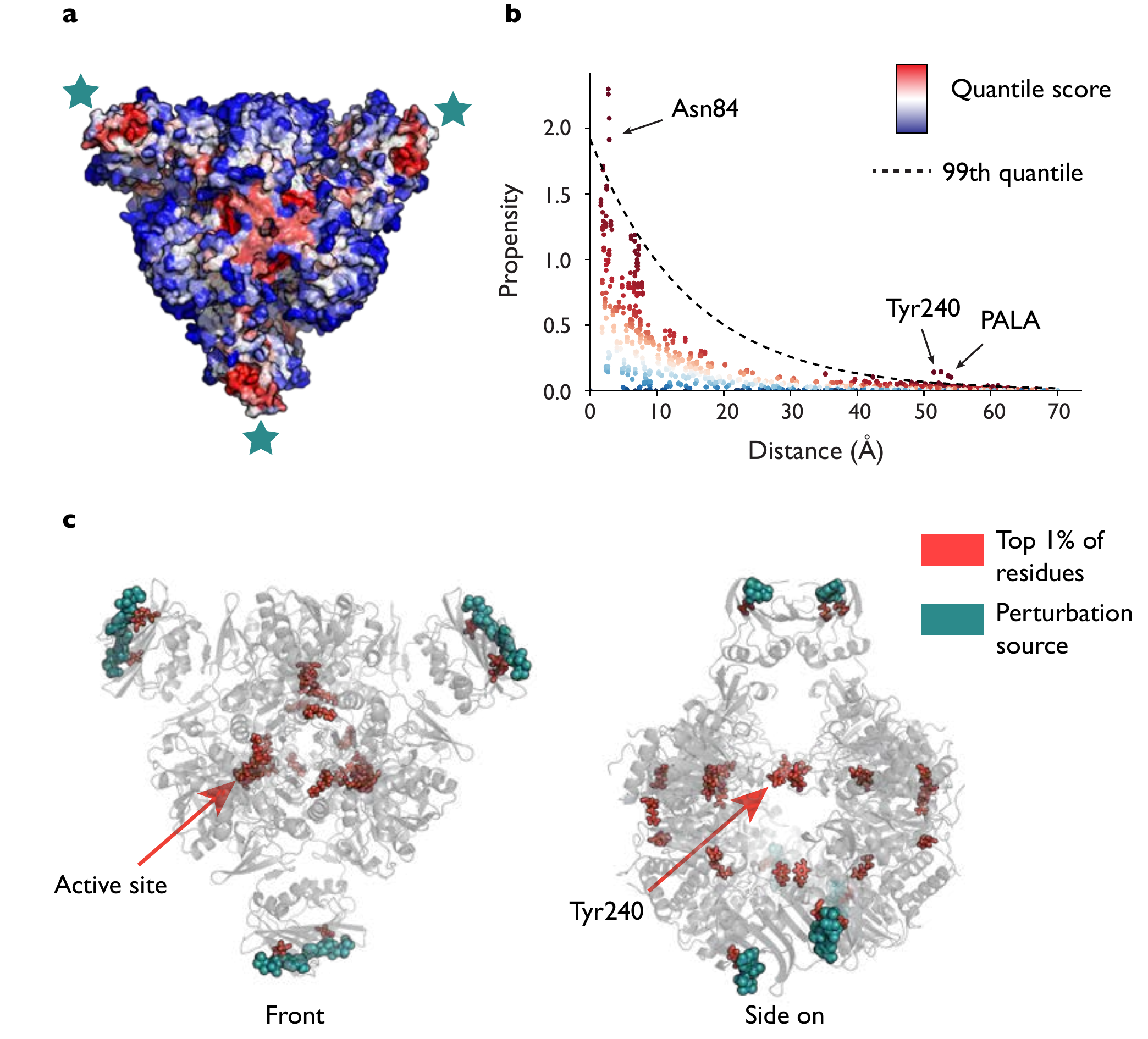}
\caption{\bf Residue ranking of the active R state of ATCase with 6 ATPs as the source by bond-to-bond propensities and conditional quantile regression.  All residues  are ranked (shown from a red to blue scale) and can be seen either directly on the structure (a) or plotted against distance from the source (b).  Here, we further focus on the top 1\% as the most significant  and plot them on the protein structure (c).  Thus (c) displays entirely equivalent results to (a) but the method allows us to highlight those residues that are \emph{particularly} important to energy distribution without making any changes to the underlying data.}
\label{R_state_6_ATP_method}
\end{figure}

We first consider the propensities with all six ATP allosteric substrates as source residues in order to identify residues and bonds that score highly and could thus be in some way significant to energy transfer in the active R state.
Figure \ref{R_state_6_ATP_method} demonstrates the output of the method.  The residue propensity for each residue in the protein is computed from their corresponding bond propensities~\eqref{residue_score} and all residues are then ranked by conditional quantile regression taking account of the distance of the residue from the source sites. Figure~\ref{R_state_6_ATP_method}a shows residues coloured according to a scale where red corresponds to those rank highly down to blue if ranked low. 
Our results show a strong link between the allosteric and active sites: all six instances of PALA, the bisubstrate analogue that sits in the active site, score highly (average $p_R = 0.996$). Indeed, it can be seen starkly from Figure \ref{R_state_6_ATP_method} that the highest scoring residues are concentrated at both the allosteric and active sites.  

In order to investigate the effect of energy flow in relation to allostery, we are also interested in the highest scoring residues according to $p_R$, the residue quantile score.
  
The highest scoring residue is Tyr240, with each of the six residues scoring $p_R = 1$ (Table \ref{4KGV_state_table}).  Tyr240 is known to play an important role in the $T \leftrightarrow R$ transition: each pair of tyrosine residues forms bonds between their phenyl rings in the R state across the gap between the two catalytic trimers (Fig.~ \ref{R_state_6_ATP_method}c), as opposed to an hydrogen bond to Asp271 in the T state~\cite{Kantrowitz1989, Burley23}.  In fact, Cherfils \textit{et al}~\cite{Kantrowitz1989} used site directed mutagenesis to substitute Tyr240 for phenylalanine, which has the effect of removing the hydroxyl group that forms the hydrogen bond in the T state, and the resulting mutated enzyme shifted strongly towards the R state upon addition of ATP in contrast to the wild-type protein.

\begin{figure}[!htb]
\centering
\includegraphics[width=0.8\textwidth]{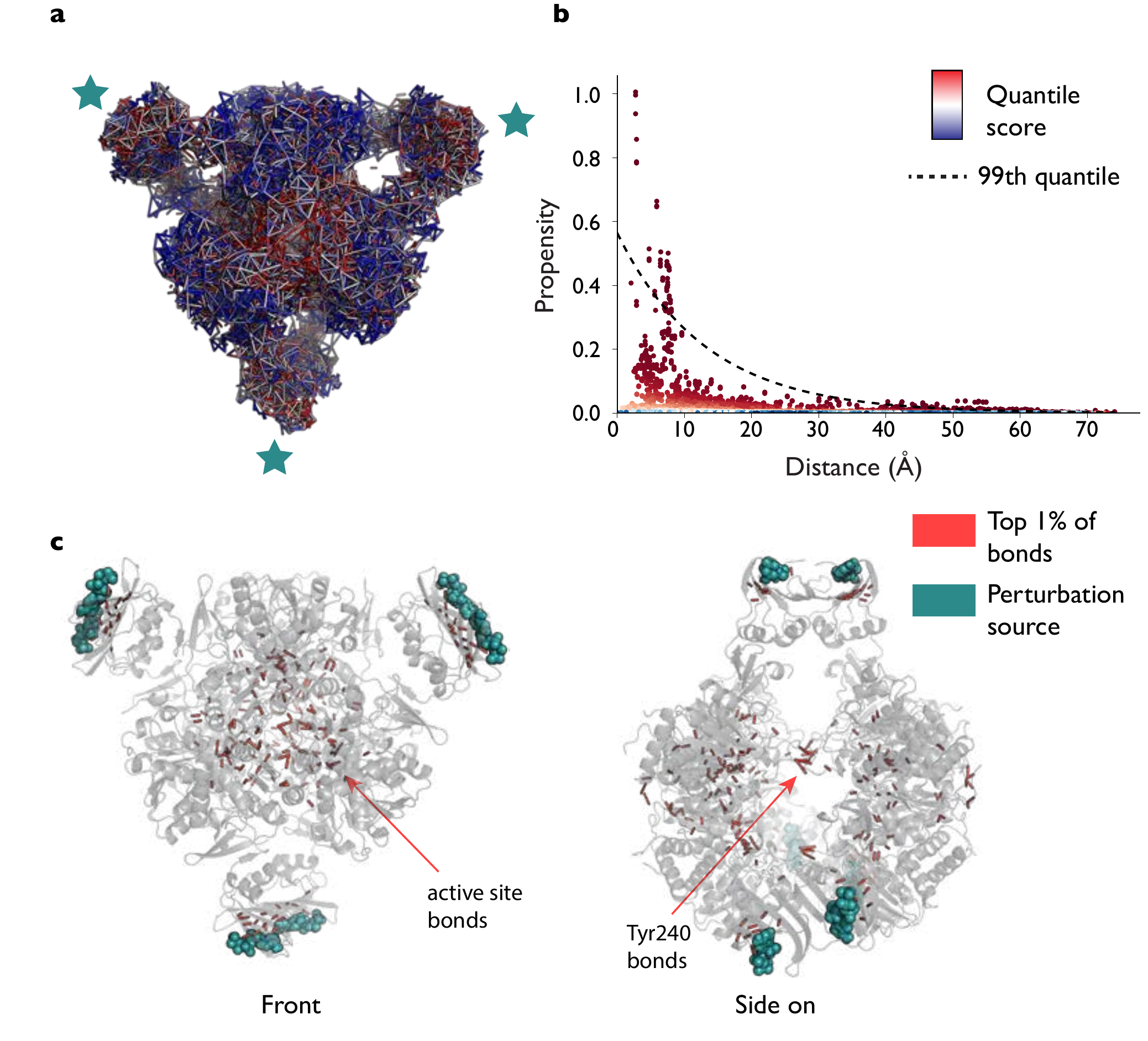}
\caption{\bf Bonds ranking in the active R state of ATCase with 6 ATPs as the source by bond-to-bond propensities and conditional quantile regression.   Each bond receives a propensity score, which is then ranked by conditional quantile regression, (a) and (b).  We can clearly highlight the highest scoring bonds by only selecting those that have scored above the $99^{th}$ percentile and  display  those bonds that are disproportionately affected by the perturbation at the six allosteric sites (c).}
\label{R_state_6_ATP_bonds}
\end{figure}

Our analysis also provides detailed bond information and we now turn to considering the bond scores directly (Fig.~\ref{R_state_6_ATP_bonds}), as key bonds within residues may be missed if other low scoring bonds in the residue 'average out' the overall residue score. One of the key bonds that emerges as significant is the hydrogen bond between Lys164 and Glu239, a bond that forms in the R state but is not present in the T state~\cite{Kantrowitz2012} (a different Lys164--Glu239 interaction exists in the T state). All six instances of this bond (from each of the six catalytic subunits) score very highly (average score of $p_b = 0.997$, where $p_b$ is the bond quantile score).  In fact, it has been shown experimentatally that when either of Lys164 or Glu239 is substituted with glutamine and lysine respectively, the mutant ATCase protein exists in the R state even in the absence of PALA and does not exhibit cooperative or allosteric effects~\cite{Newell1990}, highlighting the importance of this interaction.  
Similarly, Asn111 in the regulatory chain forms a bond with Glu109 in the catalytic chain when in the R state, and again the six instances of this hydrogen bond all score very highly. Interestingly, however, there is a slight asymmetry across the two trimers.  In chains C, G and K (See Figure \ref{ATCase_cartoon}), the average bond score is $p_b = 0.997$, whilst it is slightly lower for chains A, E and I on the other catalytic trimer ($p_b = 0.985$).  Experimental mutation of Asn111 to alanine also leads to the absence of homotropic and heterotropic effects and a shift to the R state~\cite{Eisenstein1989}.  
Another interdomain interaction identified as being highly important for stabilisation of the active R state is the Glu50--Arg234~\cite{Stieglitz2004} interaction, and we find that two different hydrogen bonds score very highly (0.995 for one set of six hydrogen bonds and 0.994 for the other) across all six catalytic chains, suggesting that the link between these two residues is particularly important for energy transfer.
This approach highlights the importance of modelling proteins at the bond level, as even course-graining to the residue level may remove crucial information. 
   
\begin{table}[!htb]
\begin{center}
 \begin{tabular}{| c  c | c c |} 
 \hline
 Residue Name and Chain & Quantile Score & Residue Name and Chain & Quantile Score \\ [0.5ex] 
 \hline
Tyr240 A & 1 & Arg105 C & 0.992\\ 
Tyr240 I & 1 & Arg105 K & 0.992\\
Pala401 C & 1 & Asp19 B & 0.992\\
Tyr240 E & 1 & Arg65 I & 0.992\\
Tyr240 C & 1 & Arg65 E & 0.992\\
Tyr240 K & 1 & Asp19 J & 0.992\\
Tyr240 G & 1 & Arg65 A & 0.992\\
Asn84 B & 1 & Asn84 H & 0.992\\
Asn84 F & 1 & Arg105 G & 0.992\\
Pala401 K & 0.996 & Arg56 K & 0.988\\
Pala401 G & 0.996 & Glu109 C & 0.988\\
Pala401 A & 0.996 & Asp19 F & 0.988\\
Arg65 C & 0.996 & Arg56 G & 0.988\\
Arg65 K & 0.996 & Glu239 I & 0.988\\
Pala401 E & 0.996 & Arg56 A & 0.988\\
Pala401 I & 0.996 & Arg56 C & 0.988\\
Asn84 J & 0.996 & Glu109 G & 0.988\\
Asn84 L & 0.996 & Glu109 K & 0.988\\
Asn84 D & 0.996 & Val83 J & 0.988\\
Arg65 G & 0.992 & Glu50 C & 0.988\\
\hline
\end{tabular}
\end{center}
\caption{\bf The top 40 residues by \emph{quantile score} (as defined in Eq. ~\eqref{residue_QR}) in the active R state (4KGV) with six ATP sources (Fig.~ \ref{R_state_6_ATP_method}).  All six active site substrate PALA residues and all six Tyr240 residues score above the 99.5\% quantile.} \label{4KGV_state_table}
\end{table}

\subsubsection{Switching of the allosteric effect is magnified by three  ATP sources in cyclic formation}

\begin{figure}[!htb]
\centering
\includegraphics[width=0.8\textwidth]{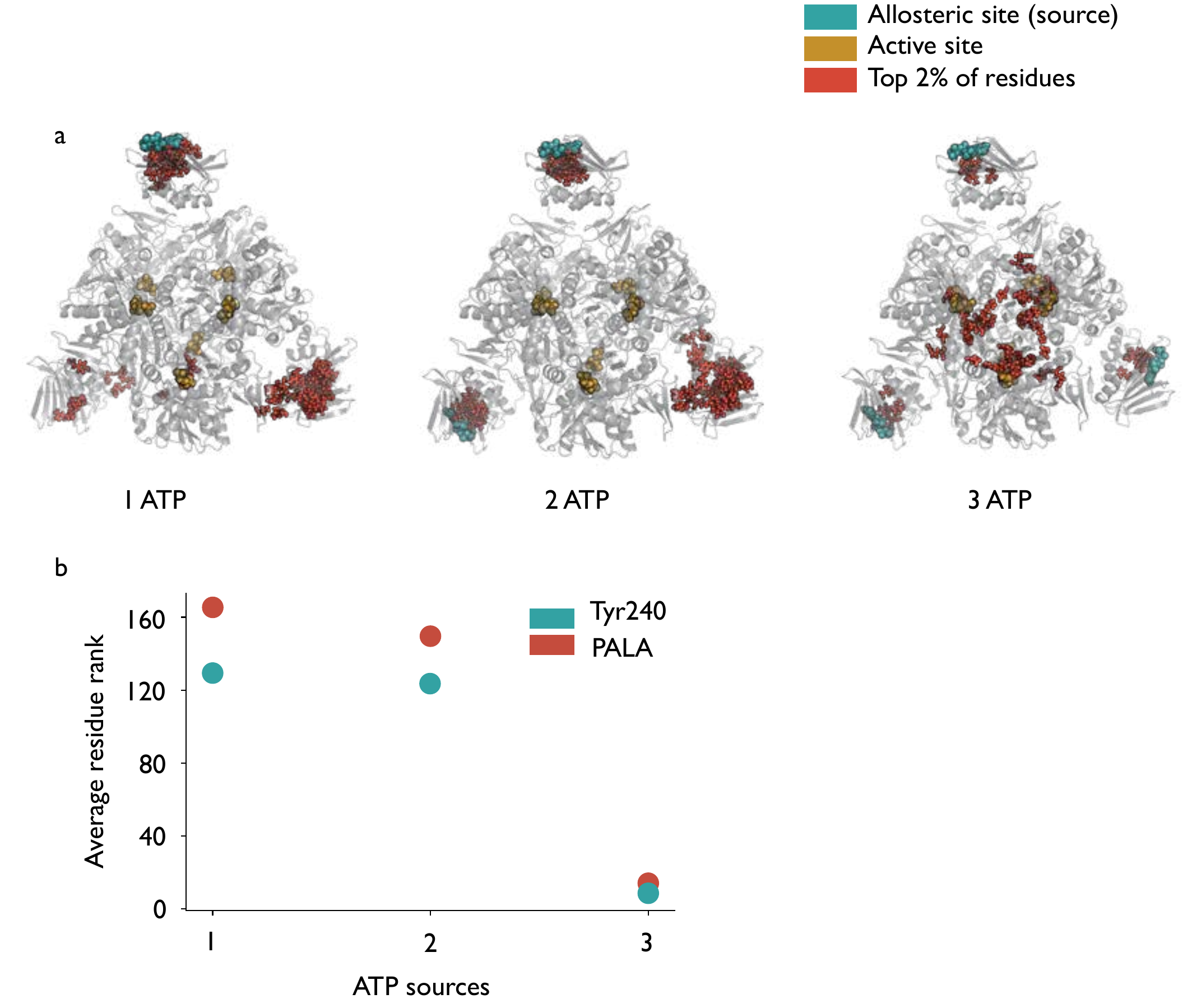}
\caption{\bf a) Binding of the first two ATP molecules does not appear to show communication between the allosteric source sites and the active site (identified by the gold PALA residues) in the active R state by bond-to-bond propensities.  However, binding of a third ATP ligand leads to a switching effect, at which point all six active site PALA residues score within the top 20 residues out of 2790. b) Scatter plot showing the average rank of the two highest scoring residues (out of 2790) from the 6 ATP case.}
\label{R_state_ATP_switch}
\end{figure}

In the previous section, all six ATP molecules were used as the source of the perturbation. The computational efficiency of our method allows us to carry out \textit{in silico} computations with different numbers of allosteric ATP molecules as perturbation sources.  We can then model the effect of progressively adding more ATP molecules to ATCase to investigate how energy flow is modified when different numbers of ligand sites are occupied. Fig.\ref{R_state_ATP_switch} shows the results of this analysis, starting with a single ATP source on chain B and adding further ligands on chain F followed by  chain J. 

When a single ATP source in regulatory chain B is used, the ranking of the residues in the protein is starkly different to the result when all six ATP molecules are used as the source (compare Figs.~\ref{R_state_6_ATP_method} and \ref{R_state_ATP_switch}).  For example, the active site residue PALA decreases its average quantile score from 0.996 (with 6 ATP molecules) to 0.941 (with 1 ATP molecule).
Instead, most of the highest scoring residues when 1 ATP is present are located near the allosteric site on chain L, which is situated across the chain B source in the multimer (Fig \ref{ATCase_cartoon}).  For instance, Asp19 (which binds to ATP) on chain L scores $p_R = 1$ and Lys56 scores $p_R = 0.996$.  Indeed, experimental mutation of Lys56 to alanine led to the disappearance of homotropic cooperativity in the presence of ATP, but not CTP~\cite{Corder1989}, suggesting it is involved in the communication pathway between ATP and the active site. 
It is also interesting to consider Tyr240, the other highly significant residue in the case of full allosteric occupation with six ATP sources.  Under single ATP occupation, the pair of Tyr240 residues in chains E and K still score highly ($p_R = 0.993$ and 1 respectively) whilst the Tyr240 residues in the other four catalytic chains score lower (average of 0.932 across the four catalytic chains).  As Fig \ref{ATCase_cartoon} shows, catalytic chains E and K are in fact situated on the other side of the protein to the chain B source, again suggesting that communication within ATCase is long range. 
When a second ATP molecule (on chain F) is included in the perturbation source, the results are similar with significant residues appearing again in the region of the allosteric sites on chains J and L distal to both source sites on chains B and F.  The PALA score on the active site is again similar to the single ATP source case ($p_R = 0.946$) showing little change in the communication to the active site upon `binding' of a second ligand.  Tyr240 scores slightly higher on average here ($p_R = 0.955$), though no single Tyr240 residues scores as high as in the six ATP case.  Overall, there is no significant change in the propensities in the active and allosteric sites to source perturbations when comparing between one or two ATP molecules binding.  

In contrast, a significant change occurs upon addition of a third ATP molecule in chain J to the perturbation source, as seen in Figure~\ref{R_state_ATP_switch}.  The average score for PALA now jumps to $p_R = 0.996$, the same as in the six ATP case, whilst Tyr240 achieves a score of $p_R = 0.998$.  Importantly, if the third ATP source is added instead to chain D (see Fig.~\ref{ATCase_cartoon}), the increases are not as pronounced (PALA = 0.948 and Tyr240 = 0.962), suggesting that the symmetric distribution of the ATP sources introducing cycles in the protein may be important for facilitating communication with the active site.

\subsection{Cooperativity: Active unligated R state with PALA sources at the active sites}

\begin{figure}[!htb]
\centering
\includegraphics[width=0.8\textwidth]{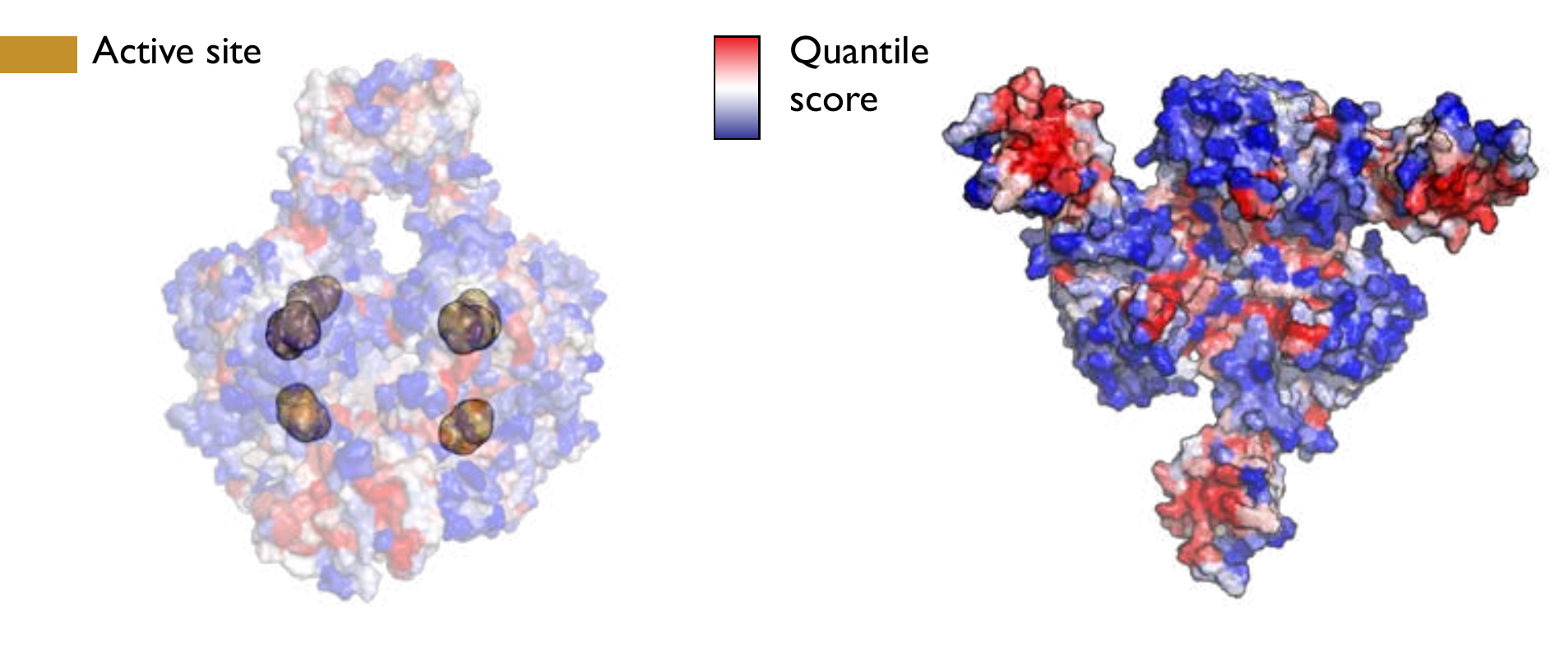}
\caption{\bf To investigate homotropic cooperativity, the six PALA substrates were selected as the source on the active unligated R state.  The structure on the right shows just one half on the protein for clarity and here it is clear that the highest scoring regions (in red) of ATCase are located around the active and allosteric sites.}
\label{R_state_No_ligand_1d09}
\end{figure}

To investigate energy flow in relation to homotropic cooperativity we analyse the full atomistic graph obtained from the crystal structure of ATCase in the active R state (1D09) with PALA molecules bound to the active sites. PALA acts as a bisubstrate analogue and, as previously, all six PALA residues are included as source residues in order to identify residues that are significant with respect to energy distribution and thus may be implicated in the cooperative mechanism.

Figure \ref{R_state_No_ligand_1d09} shows the overall effect of a perturbation applied at the six active sites.  Similarly to the `reverse' process when ATP is used as the source (compare to Fig.~\ref{R_state_6_ATP_method}), the highest scoring regions are clustered around both the allosteric and active sites.  The result reinforces the idea that there is a form of communication between these distal sites and there do not appear to be obvious, individual pathways between the two types of site.

As seen in Table \ref{1D09_state_table}, the highest scoring residue is Glu50, with all six instances scoring the maximum of $p_R = 1$.  As mentioned earlier, Glu50 is a crucial residue for the stability of the R state: substitution of glutamic acid for alanine leads to dramatic changes in the activity of the enzyme, which is reduced 15-fold, whereas cooperativity is completely lost~\cite{Newton1990}.  
Significant communication from the active sites to the allosteric sites is also seen, with Asp19 (one of the residues that interacts with ATP and CTP) scoring $p_R = 0.992$ over the six sites, whilst Lys60, another allosteric residue, scores highly ($p_R = 0.989$) over the regulatory chains on one side of the protein (chains D, H and L in Fig.~\ref{ATCase_cartoon}), again demonstrating asymmetry over the structure.   
It is known experimentally that Glu233 forms a salt link with Arg229 only in the R state, which orients Arg229 into the active site~\cite{Ke1988}.  The removal of the salt link via mutation of glutamic acid to serine leads a significant decrease in both catalytic activity and cooperativity~\cite{Biol1989}. We find that Glu233 scores highly overall ($p_R = 0.985$), though once again a difference is seen between the two catalytic trimers (trimer AEI scores $p_R = 0.993$ vs $p_b = 0.977$ for the opposite trimer CGK).

Analysis of the \emph{bond level} data reveals further information.  As expected, the previously mentioned Glu233--Arg229 salt bridge ranks very highly ($p_b = 0.996$), whilst the Glu50 interaction with Arg167 (which itself interacts with PALA in the active site, being positioned correctly by its association with Glu50) involving two types of bonds scores above $p_b = 0.995$ across all units.  The Asp19--Lys56 link scores an average of $p_b = 0.999$ over its six instances and it was found that substitution of lysine by alanine affected not only cooperativity but also removed the ability of ATP to activate the enzyme~\cite{Corder1989}.  Therefore, as Asp19 is one of the allosteric residues, it appears that this bond to Lys56 may be crucial in communicating with the active site.

\begin{table}[!htb]
\begin{center}
 \begin{tabular}{| c  c | c c |} 
 \hline
 Residue Name and Chain & Quantile Score & Residue Name and Chain & Quantile Score \\ [0.5ex] 
 \hline
Asp19 D & 1 & Lys60 H & 0.993\\ 
Asp19 H & 1 & Leu58 F & 0.993\\
Asp19 L & 1 & Ile18 D & 0.993\\
Glu50 E & 1 & Ile18 H & 0.993\\
Glu50 A & 1 & Asp90 K & 0.993\\
Glu50 K & 1 & Asp90 I & 0.993\\
Glu50 G & 1 & Glu233 I & 0.993\\
Glu50 I & 1 & Glu233 E & 0.993\\
Glu50 C & 1 & Glu233 A & 0.993\\
Met1 H & 0.996 & Arg167 E & 0.993\\
Ile44 L & 0.996 & Asp19 F & 0.989\\
Ile44 H & 0.996 & Lys60 D & 0.989\\
Ile44 D & 0.996 & Leu58 B & 0.989\\
Asp90 E & 0.996 & Leu58 J & 0.989\\
Asp90 A & 0.996 & Asp90 G & 0.989\\
Asp90 C & 0.996 & Arg269 A & 0.989\\
Arg105 G & 0.996 & Glu86 E & 0.989\\
Arg105 C & 0.996 & Arg167 I & 0.989\\
Arg105 K & 0.996 & Arg167 G & 0.989\\
Met1 D & 0.993 & Arg167 K & 0.989\\
\hline
\end{tabular}
\end{center}
\caption{\bf The top 40 residues by \emph{quantile score} in the active R state (1D09) with six PALA sources (see Fig. \ref{R_state_No_ligand_1d09}).  Each of the six Glu50 residues score the maximum value of 1.} \label{1D09_state_table}
\end{table}

\subsection{Stabilisation of the catalytic trimer: inactive T state with allosteric CTP}

\begin{figure}[!htb]
\centering
\includegraphics[width=0.8\textwidth]{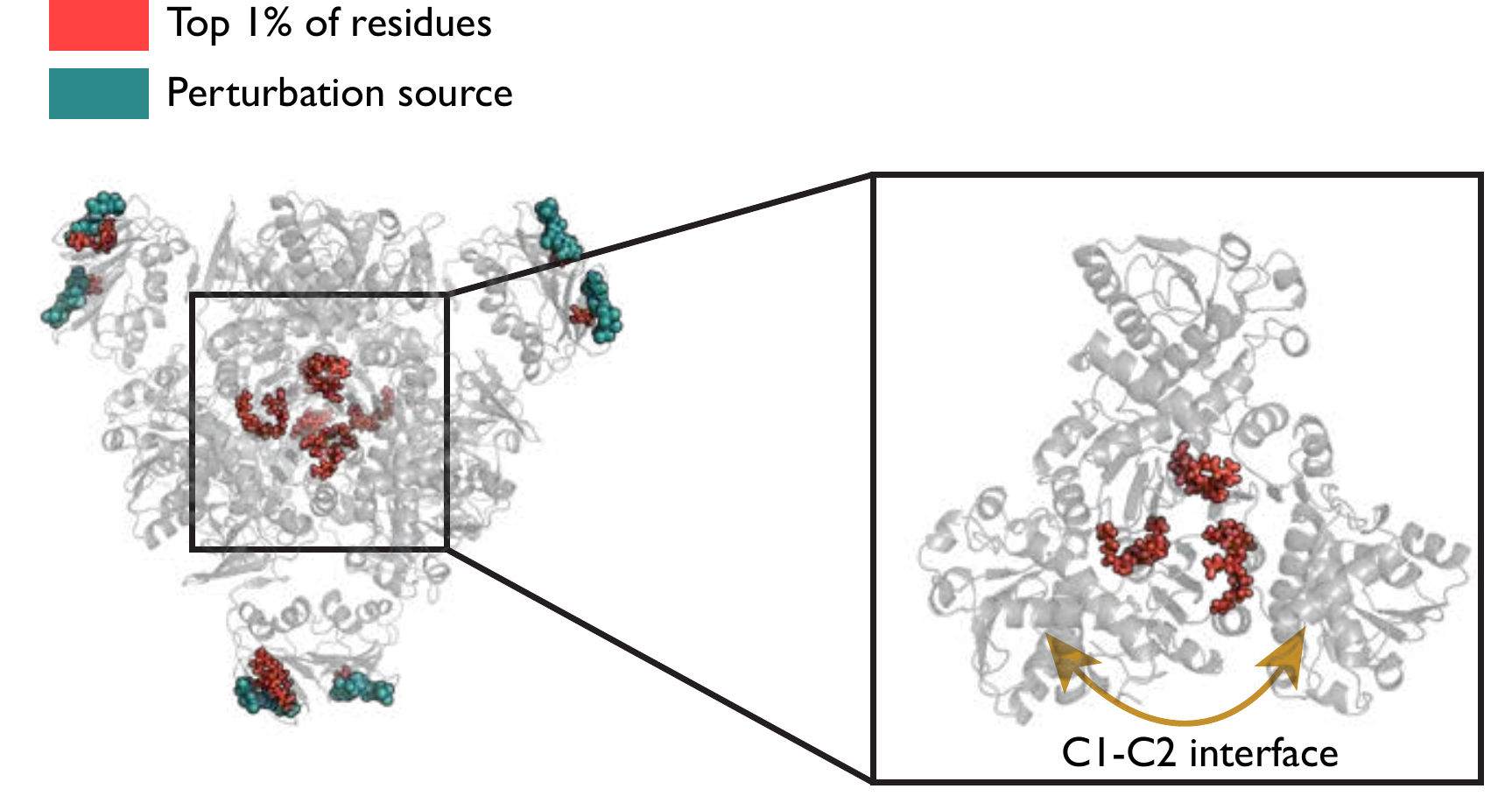}
\caption{\bf When six CTP molecules are used as the perturbation source on the inactive T state, the highest scoring residues appear at the C1-C2 interface, which is the boundary between catalytic subunits within the catalytic trimer.  Arg56 and Arg65 are two of the highest scoring residues,  shown on the right, forming a link across the C1-C2 interface.}
\label{T_state_CTP_6_source}
\end{figure}

Finally, we study the inactive state through the analysis of the graph obtained from the crystal structure of ATCase in the T state (5AT1) with CTP molecules bound to the allosteric sites.  CTP acts as an inhibitor  for ATCase reducing its catalytic rate by causing a shift towards the inactive T state.  When using the full allosteric occupation scenario (6 CTP molecules as perturbation sources) in the T state,
the highest scoring regions of the protein appear most strongly located at the C1--C2 interface (Fig.~\ref{T_state_CTP_6_source}) instead of the active site.  This is in stark contrast with the results for the R state, both under ATP (allosteric site) and PALA (active site) perturbation sources, as shown in Figs.~\ref{R_state_6_ATP_method}~and~\ref{R_state_No_ligand_1d09}.
The two catalytic trimers each move as essentially rigid units during the T $\leftrightarrow$ R transition so there is little structural change within the trimer.  There is thus only minor differences between the inactive T state and the active R state in this region~\cite{Krause1987}.   

Two residues in particular stand out, as seen in Table \ref{T_state_table}: Arg65 (average $p_R = 0.999 $) and Arg56 (average $p_R = 0.998 $).  It can be seen from Figure \ref{T_state_CTP_6_source} that both these residues bridge the C1--C2 interface, though they do not form links to each other.  
Looking in more detail at the bond level data, one of the key interactions made by Arg65 is with Asp100 (average $p_b = 0.999$).  This specific interaction was identified experimentally as being important for the stability of the catalytic trimer~\cite{Baker1993} and replacement of Asp for either Asn or Ala reduces the half life of inactivation of the catalytic subunit. 
Arg65 additionally forms a hydrogen bond to His41, another residue implicated in catalytic subunit stability and this interaction also scores highly ($p_b = 0.983$), though once again there is a significant difference between the two catalytic subunits, with the interactions in the AEI trimer scoring $p_b = 0.999$, compared to $p_b = 0.968$ in trimer CGK (Fig.~ \ref{ATCase_cartoon}).  There is possibly a link here with experimental data showing that in the R state, only half (i.e. three) of the His41--Glu37 interactions are broken~\cite{Gouaux1990, Stevens1991a} during the transition from the T state, demonstrating an intriguing asymmetry that appears to be captured by our computed \emph{bond-to-bond propensities}.  In fact, the residue results for Glu37 are even starker, with the average quantile score across chains A, E and I 0.990 versus 0.262 for chains C, G and K, a remarkable difference between essentially symmetrically equivalents sets of residues.  Glu37 itself has been associated with stabilising the catalytic trimer~\cite{Baker1993}. 

Conversely, there appears to be little experimental data on Arg56, nor on the two highest scoring links it makes: to Gly72 ($p_b = 0.999$) and Gln60 ($p_b = 0.986$), though the Gly72 interaction occurs across the C1--C2 interface~\cite{Jin1999, Stevens1991a} so it would seem possible that this interaction is also involved in stability of the trimer.  Perhaps less surprisingly, a number of residues located close to the CTP site also rank highly: Ile86, which forms a non-polar interaction with the nucleotide~\cite{Kantrowitz2012}) and Asn84, which interacts with the phosphate part of CTP~\cite{Stevens1991a} score $p_R = 0.993$ and $0.985$ respectively.  Val17 also forms a non-polar interaction with CTP, though scores slightly lower with an average bond quantile score of $0.978$.

\begin{table}[!htb]
\begin{center}
 \begin{tabular}{|c  c | c c|} 
 \hline
 Residue Name and Chain & Quantile Score & Residue Name and Chain & Quantile Score\\ [0.5ex] 
 \hline
Arg65 G & 1 & Val71 G & 0.993\\
Arg56 G & 1 & Val71 A & 0.993\\
Arg65 C & 1 & Val71 K & 0.993\\
Arg65 A & 1 & Glu37 A & 0.993\\ 
Arg56 A & 1 & Val71 C & 0.993\\
Arg65 K & 1 & Gln60 E & 0.993\\
Asn84 H & 1 & Gln60 A & 0.993\\
Asn84 D & 1 & Val83 H & 0.993\\
Ile86 J & 1 & Val17 H & 0.993\\
Arg56 K & 0.996 & Val17 D & 0.993\\
Arg65 I & 0.996 & Ile86 D & 0.993\\
Arg65 E & 0.996 & Arg105 A & 0.989\\ 
Arg56 C & 0.996 & Gln60 I & 0.989\\
Arg56 I & 0.996 & Phe73 K & 0.989\\
Arg56 E & 0.996 & Glu37 E & 0.989\\
Arg85 H & 0.996 & Arg105 E & 0.989\\
Ile86 F & 0.996 & Glu37 I & 0.989\\
Ile86 B & 0.996 & Arg85 D & 0.989\\
Asn84 L & 0.996 & Val83 D & 0.989\\
Ile86 H & 0.996 & Ile12 J & 0.989\\
\hline
 \end{tabular}
\end{center}
\caption{\bf T state with CTP (5AT1).  The top 40 residues by quantile score are listed and both Arg56 and Arg65 appear six times each.  These residues sit at the C1-C2 interface within the catalytic subunits.}
\label{T_state_table}
\end{table}

\subsubsection{Sequential binding of CTP and ATP show similar switching patterns but on different regions of the protein}

\begin{figure}[!htb]
\centering
\includegraphics[width=0.8\textwidth]{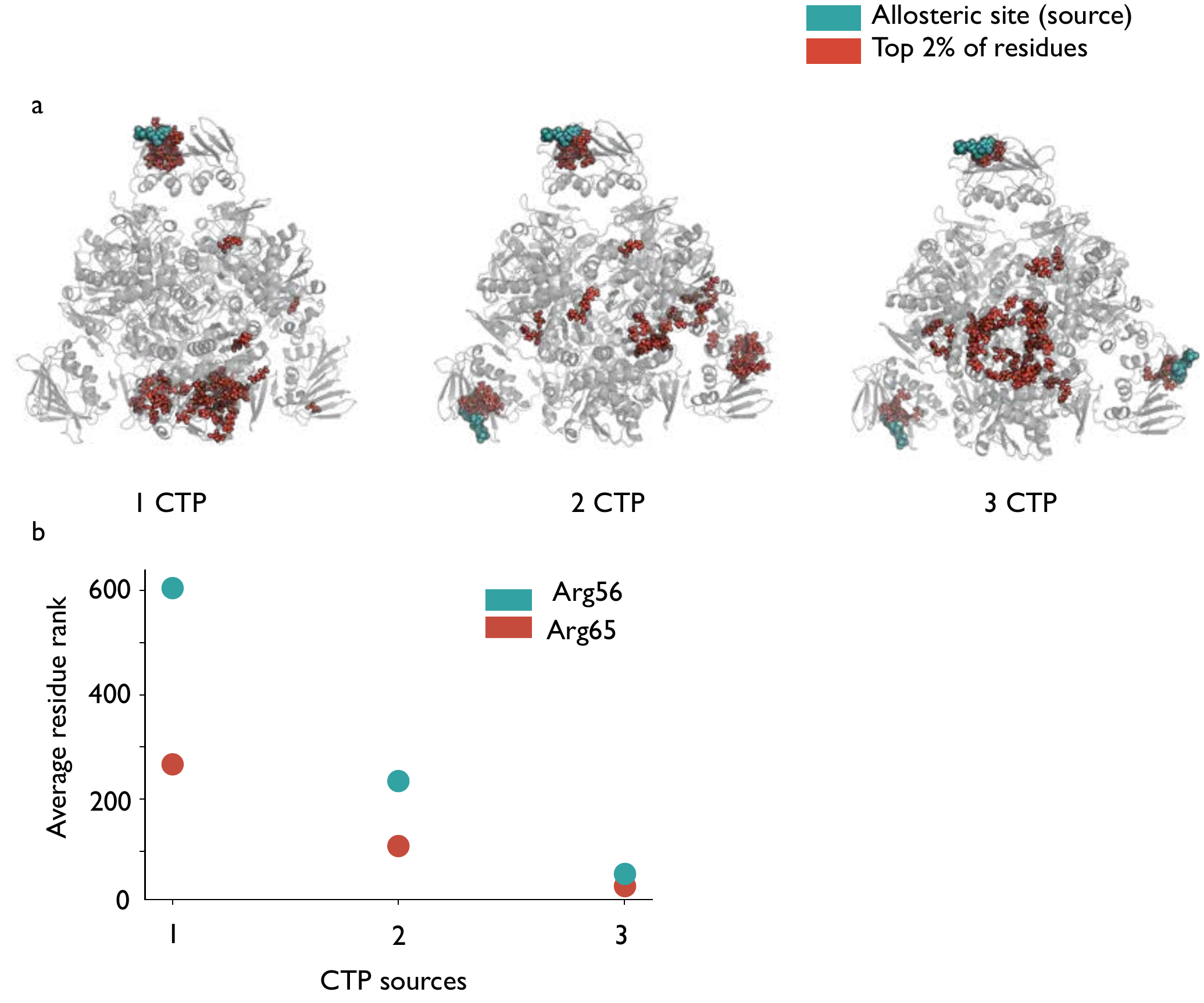}
\caption{\bf a) The top 2\% of residues displayed when varying numbers of CTP molecules are included as source residues.  In contrast to the ATP case, there does not appear to be as much communication with the distal allosteric sites for one or two source ligands but again inclusion of a third ligand on chain J leads to the results resembling the six CTP case described previously. b) Scatter plot showing the average rank of the two highest scoring residues (out of 2790) from the 6 CTP case.}
\label{T_state_CTP_switch}
\end{figure}

Whilst the identity of the highest scoring residues when CTP is used as the perturbation source is different to the ones that appeared when ATP is used, there is a similar 'switching effect' observed when a third CTP molecule is included as a source in a cyclic arrangement around the ATCase structure.  As seen in Figure \ref{T_state_CTP_switch}, inclusion of a third ligand leads to the clustering of high scoring residues in the region of C1--C2 interface between the catalytic subunits within a trimer, leading to a similar pattern observed under the full occupation six CTP source case (Fig.~\ref{T_state_CTP_6_source}).  

Once again, it appears to be the interaction between the CTP ligands located in such a symmetric, cyclic arrangement around the ATCase protein that leads to energy flow amplifying the effect on residues identified as particularly significant.
This non linear effect is similar to the one observed for ATP allosteric occupation in Figure~\ref{R_state_ATP_switch}, albeit in a different location in the protein.
This effect is illustrated numerically by focusing on two of the highest scoring residues: Arg56 and Arg65.  Starting from a single CTP source, the scores for Arg65 progress from $p_R = 0.904$ to 0.961 and then to 0.989 when a third ligand is included; equivalently, for Arg56, the scores are 0.779, 0.916 and 0.982 as each of the CTP ligands is added.  
The increases in scores of these two highest scoring residues in this case are actually more linear than in the case of ATP but it is still only when a third CTP ligand is included cyclically that the results from the six CTP case are replicated.  
When the third ligand is instead added to chain D, such that the three CTP ligands are now bound to chains B, D and F (Fig.~\ref{ATCase_cartoon}), the increase in score upon addition of the third ligand is smaller for both Arg65 ($p_R = 0.973$) and Arg56 ($p_R = 0.922$) which again suggests that it is a particular feature of the symmetric arrangement of the allosteric ligands that facilitates communication to the key residues within the protein by creating cyclic reinforcement of energy flows in the protein.

\section{Discussion}

In this work, we have demonstrated how \emph{bond-to-bond propensities} can be used to investigate the energy flow process of a heterotropic ligand binding to an allosteric site, as well as the homotropic case of substrates binding to the active site.  We have focused on a large (\~43,000 atom) well studied multimeric enzyme, ATCase which shows both allostery and cooperativity. In the active R state, using ATP allosteric binding sites as perturbation sources, reveals a number of residues and specific bonds as being particularly significant, including Tyr240, which links the two sides of the ATCase protein, and PALA, which sits in the active site.  There is thus a clear communication pathway between the allosteric and active sites in ATCase but in accordance with other computational studies of ATCase~\cite{Mitchell2016}, this communication does not appear to occur through individual, discrete pathways of residues but instead via a collective of lower scoring residues.

Furthermore, we find that the geometrical distribution of the ligands bound is important leading to a switching of the allosteric effect in our computations.  Only when three ATP residues arranged cyclically around the ATCase structure are used as the perturbation source do we recover the results of the case with full allosteric site occupation. In contrast, when a single or two ATP molecules are used as sources, no strong link between the allosteric and active sites is observed, though there does seem to be communication between distal allosteric sites. A feature of our results is a consistent asymmetry in the scores between the two sets of catalytic trimers, despite the symmetrical structure of ATCase, demonstrating the ability of \emph{bond-to-bond propensities} to capture subtle structural features at the atomistic level.

Homotropic cooperativity was investigated by using the six PALA substrates bound at the active sites as the perturbation source.  The regions that scored most highly in this case were the allosteric sites and around the bound active site, reinforcing the idea that the two types of sites are highly coupled in the active state and also hinting that homotropic and heterotropic cooperative are not orthogonal phenomena and are, instead, closely intertwined. 

Finally, allosteric inhibition of ATCase by CTP was studied by using the CTP molecules as the perturbation source.  Interestingly, rather than the active site region being identified as significant,  the C1--C2 interface of the catalytic trimers was instead found to be signficantly coupled to the allosteric sites.  Interestingly, the boundary between the catalytic subunits has been found to be important for stability of the enzyme but not particularly vital for catalytic activity; hence it is possible that different allosteric ligands may play subtly different roles when binding to the active and inactive states of the enzyme.

Our results highlight that both the atomistic nature of the methodology and the long-range effects made possible by the global properties of the graph-theoretical approach are essential to understanding the effects of allostery and cooperativity in multimeric protein. Given its computational efficiency and generality we hope that this method can be useful to the study of other such protein systems of broad relevance.

\section*{Acknowledgments}

We thank Ben Amor for useful discussions.  This work was funded by an EPSRC Centre for Doctoral Training Studentship from the Institute of Chemical Biology (Imperial College London) awarded to MH. MB acknowledges funding from the EPSRC project EP/N014529/1 supporting the EPSRC Centre for Mathematics of Precision Healthcare.

\section*{Contributions}
M.H., S.N.Y. and M.B. designed the experiments, M.H. carried out the experiments and analyzed data, M.H., S.N.Y. and M.B. wrote and approved the manuscript.

\section*{Competing interests}
The authors declare no competing interests.

\section*{Corresponding author}
Correspondence to Sophia N. Yaliraki (s.yaliraki@imperial.ac.uk).

\bibliographystyle{naturemag}
\small\bibliography{ATCase_paper}

\end{document}